\documentstyle[aps,eqsecnum,epsf,preprint]{revtex}
\input epsf.tex
\def\.{\!\cdot\!}
\def\:{\cdots}
\def\[{\left[}
\def\]{\right]}
\def\({\left(}
\def\){\right)}

\def\bi{\begin{itemize}}

\def\ei{\end{itemize}}
\def\be{\begin{eqnarray}}
\def\ee{\end{eqnarray}}
\def\bn{\begin{enumerate}}
\def\en{\end{enumerate}}
\def\h{{1\over 2}}

\def\nn{\nonumber\\}

\def\r2{\sqrt{2}}

\def\P{{\cal P}}
\def\o{\omega}
\def\e{\epsilon}

\def\t{\tilde}
\def\L{\Lambda}

\def\db{\delta_{-1}}
\def\da{\delta_0}

\begin{document}
\title{Decomposition of 
Time-Ordered Products and Path-Ordered Exponentials}
\author{C.S. Lam\cite{email}}
\address{Department of Physics, McGill University,
 3600 University St., Montreal, QC, Canada H3A 2T8}
\maketitle

\begin{abstract}
We present a decomposition formula for $U_n$,
an integral of time-ordered
products of operators, in terms of sums of products of the more primitive 
quantities $C_m$, which are the
integrals of time-ordered commutators of 
the same operators.  
The resulting factorization enables a summation over $n$ to be carried out
to yield an explicit expression for the time-ordered exponential, an 
expression which turns out to be an 
exponential function of
$C_m$.
The Campbell-Baker-Hausdorff formula and the nonabelian eikonal formula
obtained previously are both special cases of this result.
\end{abstract}

\section{Introduction}

The path-ordered exponential $U(T,T')= \P\exp(\int_{T'}^TH(t)dt)$
is the solution of the first order differential equation 
$dU(T,T')/dT=H(T)U(T,T')$ with the initial condition $U(T',T')={\bf 1}$.
The function $H(T)$ can be
 a finite-dimensional matrix or an infinite-dimensional
operator. In the latter case all concerns of domain 
and convergence will be ignored.
The parameter $t$ labelling the path shall be
referred to as `time', so path-ordering and time-ordering
are synonymous in the present context.  
The path-ordered exponential is usually computed from its
power series expansion, $U(T,T')=\sum_{n=0}^\infty U_n$,
 in terms of the
time-ordered products $U_n=\P(\int_{T'}^T H(t)dt)^n/n!$. 

Path-ordered exponential
 can be found in many areas of theoretical physics. 
It is the time-evolution
operator in quantum mechanics if $iH$ is the Hamiltonian. It is the
non-integrable phase factor (Wilson line) of the Yang-Mills theory if 
$-iH=A$ is the nonabelian vector potential. It defines  
an element of a Lie group connected to the identity via
 a path whose tangent
vector at time $t$ is given by the Lie-algebra element $-iH(t)$.
It can be used to discuss canonical 
transformation and classical particle trajectories \cite{dragt}.
It also gives rise to useful formulas in many
other areas of mathematical physics
 by  suitable choices of $H(t)$, some of which
 will be mentioned later. Likewise, time-ordered products are
ubiquitous. For example, free-field
matrix elements of time-ordered
products of an interaction Hamiltonian give rise to perturbation diagrams.

The main result of this paper is a 
{\it decomposition theorem} for the time-ordered products $U_n$.
It provides a formula equating them to
sums of products of the more primitive quantities $C_m$. 
The factorization thus achieved enables $U_n$ to be summed
up and $U(T,T')$ expressed as an exponential function of the $C_m$'s.

This result grew out of a nonabelian eikonal formula.
The abelian version of the eikonal formula \cite{EIK}
is well known. It gives rise to a geometrical interpretation for elastic
scattering at high energies \cite{geom}, besides
being  a useful tool in dealing with
infrared divergence \cite{IR} in QED. The nonabelian version was 
originally developed 
to deal with the consistency problem of baryonic amplitudes in
large-$N_c$ QCD \cite{LL1,LL2}, but it proves to be useful also in  
in understanding gluon reggeization and reggeon factorization
 \cite{FHL,FL}. It is expected to be valuable also in 
the discussion of
geometrical pictures and infrared problems in QCD.

The generalization in this paper can be used to take the nonabelian
eikonal formula a step further to include small transverse motions
of the high-energy trajectory. Such corrections to the usual eikonal
formula are known to be crucial in obtaining the Landau-Pomeranchuk-Migdal 
effect \cite{LPM} in QED, so are expected to be important as well in the
case of QCD. However, we shall postpone all such physical applications
and concentrate in this paper to the development of mathematical formulas.

On the mathematical side, one can derive the 
Campbell-Baker-Hausdorff formula \cite{bourbaki}
as a corollary of the formulas developed in these pages.

Statements, explanations, and simple illustrations of the 
results will
be found in the main text, while proofs and other details
will be relegated to the Appendices. 

The formulas are combinatorial in character. They cannot be adequately
explained without a suitable set of notations, which we develop in
Sec.~2. The main result of the paper, the decomposition theorem, will
be discussed in Secs.~3 and 4. This theorem can be
 stated for a more general
time-ordered product $U_n$, in which all the $H_i(t)$ are different.
This will be dealt with in Sec.~3.
The special case when these $H_i(t)=H(t)$ are identical will be taken up
in Sec.~4. This in turn leads to the exponential formula for $U(T,T')$
in Sec.~5. The final section, Sec.~6,
 is put in mainly to illustrate the versatility
of our results in deriving other formulas useful in mathematics
and physics, by suitable choices of $H_i(t)$.
Among them are the 
Campbell-Baker-Hausdorff formula for the multiplication of group
elements, and the nonabelian generalization of the eikonal formula
in physics.

\section{Definitions}

We start by generalizing the definition of $U_n$ to cases when the 
operators $H(t)$ are all different.

Let $[s]=[s_1s_2\cdots s_n]$ be a permutation of the $n$ numbers $[12\cdots n]$, and $S_n$ the corresponding permutation group. We
define the 
{\it time-ordered product} $U[s]$ to be the integral
\be
U[s]=&&U[s_1s_2\cdots s_n]=\int_{R[s]}dt_1dt_2\cdots
 dt_n\nn
&&H_{s_1}(t_{s_1})H_{s_2}(t_{s_2})\cdots H_{s_n}(t_{s_n})\label{Us}\ee
taken over the hyper-triangular region $R[s]=\{T\ge t_{s_1}\ge t_{s_2}
\ge\cdots\ge t_{s_n}\ge T'\}$, with operator $H_{s_i}(t_{s_i})$
standing to the left of $H_{s_j}(t_{s_j})$ if $t_{s_i}>t_{s_j}$.
The operator $U_n=U_n[T,T']$ is then defined to be
the average of $U[s]$ over all permutations $s\in S_n$:
\be
U_n&=&{1\over n!}\sum_{s\in S_n}U[s]\label{Un}.\ee
In the special case when all $H_i(t)=H(t)$ are identical, this $U_n$
coincides with the one in the Introduction.

The  {\it decomposition theorem}
expresses $U_n$ as sums of products of the
{\it time-ordered commutators} $C[s]=C[s_1s_2\cdots s_n]$.
These are defined
analogous to $U[s_1s_2\cdots s_n]$, but with the products of $H_i$'s
changed to nested multiple commutators:
\be
C[s]=\int_{R[s]}&&dt_1dt_2\cdots
 dt_n
[H_{s_1}(t_{s_1}),[H_{s_2}(t_{s_2}),[\cdots ,\nn
&&[H_{s_{n-1}}(t_{s_{n-1}}),
H_{s_n}(t_{s_n})]\cdots]]].\label{Cs}\ee
For $n=1$, we let $C[s_i]=U[s_i]$ by definition.
Similarly, the operator $C_n=C_n[T,T']$ is defined to be the average 
of $C[s]$ over all permutations $s\in S_n$:
\be
C_n&=&{1\over n!}\sum_{s\in S_n}C[s]\label{Cn}.\ee
It is convenient to use a `{\it cut}' (a vertical bar) 
to denote products of $C[\cdots]$'s.
For example, $C[31|2]\equiv C[31]C[2]$, and $C[71|564|2|3]\equiv
C[71]C[564]
C[2]C[3]$.

We are now in a position to state the main theorem. 

\section{General Decomposition Theorem}

\be
n!U_n=\sum_{s\in S_n}C[s_1\cdots|\cdots|\cdots\ \cdots s_n]\equiv
\sum_{s\in S_n}C[s]_c.
\label{thm}
\ee
A cut (vertical bar) is placed after $s_i$ if and only if
$s_j>s_i$ for all $j>i$. An element $s\in S_n$ with cuts placed
this way will be denoted by $[s]_c$. 

In view of the fundamental nature of this theorem, 
two separate proofs shall be provided for it in Appendix A.
\subsection{Examples}

For illustrative purposes 
here are explicit formulas for $n=1,2,3$, and 4:
\be
1!U_1&=&C[1]\nn
2!U_2&=&C[1|2]+C[21]\nn
3!U_3&=&C[1|2|3]+C[21|3]+C[31|2]+C[1|32]\nn
&+&C[321]+C[231]\nn
4!U_4&=&C[1|2|3|4]+C[321|4]+C[231|4]+C[421|3]\nn
&+&C[241|3]+C[431|2]
+C[341|2]
+C[1|432]\nn &+&C[1|342]+C[21|43]+C[31|42]+C[41|32]\nn
&+&C[21|3|4]
+C[31|2|4]+C[41|2|3]+C[1|32|4]\nn &+&C[1|42|3]+C[1|2|43]
+C[4321]+C[3421]\nn
&+&C[4231]+C[3241]+C[2341]+C[2431]
\label{ex1}\ee

The formula (\ref{thm}) can be displayed graphically, with a filled
circle with $n$ lines on top indicating $U_n$, and an open circle with
$n$ tops on top indicating $C[s]$. This is shown in Fig.~1 for $n=3$.

\begin{figure}
\vskip -5cm
\centerline{\epsfxsize 4.7 truein \epsfbox {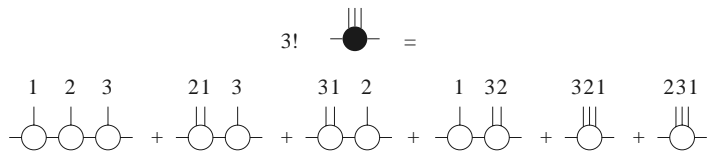}}
\nobreak
\vskip -0.5cm\nobreak
\caption{The decomposition of $3!U_3$
(filled circle) in terms of $C[s]$'s.}
\end{figure}

\subsection{Factorization}

In the special case when all the $H_i(t)$'s mutually commute,
the only surviving $C[\cdots]$'s are those with one
argument, so eq.~(\ref{thm}) reduces to a {\it factorization
theorem}:
\be
U_n={1\over n!}\prod_{i=1}^nC[i].\label{abthm}\ee
Thus the general decomposition theorem in (\ref{thm}) may be thought
of as a {
\it nonabelian generalization of this factorization 
theorem}.
\section{Special Decomposition Theorem}

Great simplification occurs when all $H_i(t)=H(t)$ are identical, 
for then $U[s]$ and $C[s]$ depend only on $n$ 
but not on the particular
$s\in S_n$. Hence all $U[s]=U_n$
and all $C[s]=C_n$. In this case the decomposition theorem becomes
\be
U_n&=&\sum\xi(m_1m_2\cdots m_k)C_{m_1}C_{m_2}\cdots C_{m_k},\nn
\xi(m)&\equiv&
\xi(m_1m_2\cdots m_k)=\prod_{i=1}^k\[\sum_{j=i}^km_j\]^{-1}.
\label{cor1}\ee
The sum in the first equation is taken over all $k$, and 
all $m_i>0$ such that $\sum_{i=1}^km_i=n$.
Note that  $\xi(m)=\xi(m_1m_2\cdots m_k)$ is 
{\it not} symmetric under the interchange of the $m_i$'s.
It is this asymmetry that produces the commutator terms
in the  formulas for $K_n$
in eq.~(\ref{exp}).

See Appendix B for a proof of this special decomposition theorem.

We list below this special decompositions up to $n=5$:
\be
1!U_1&=&C_1\nn
2!U_2&=&C_1^2+C_2\nn
3!U_3&=&C_1^3+2C_2C_1+C_1C_2+2C_3\nn
4!U_4&=&C_1^4+6C_3C_1+2C_1C_3+3C_2^2+3C_2C_1^2\nn
&+&2C_1C_2C_1+C_1^2C_2
+6C_4\nn
5!U_5&=&C_1^5+24C_4C_1+6C_1C_4+12C_3C_2+8C_2C_3\nn
&+&12C_3C_1^2
+6C_1C_3C_1
+2C_1^2C_3+8C_2^2C_1\nn
&+&4C_2C_1C_2
+3C_1C_2^2+4C_2C_1^3+3C_1C_2C_1^2\nn
&+&2C_1^2C_2C_1+C_1^3C_2.
\label{ex2}
\ee
The case for $n=3$ is explicitly shown in Fig.~2.

\begin{figure}
\vskip -5cm
\centerline{\epsfxsize 4.7 truein \epsfbox {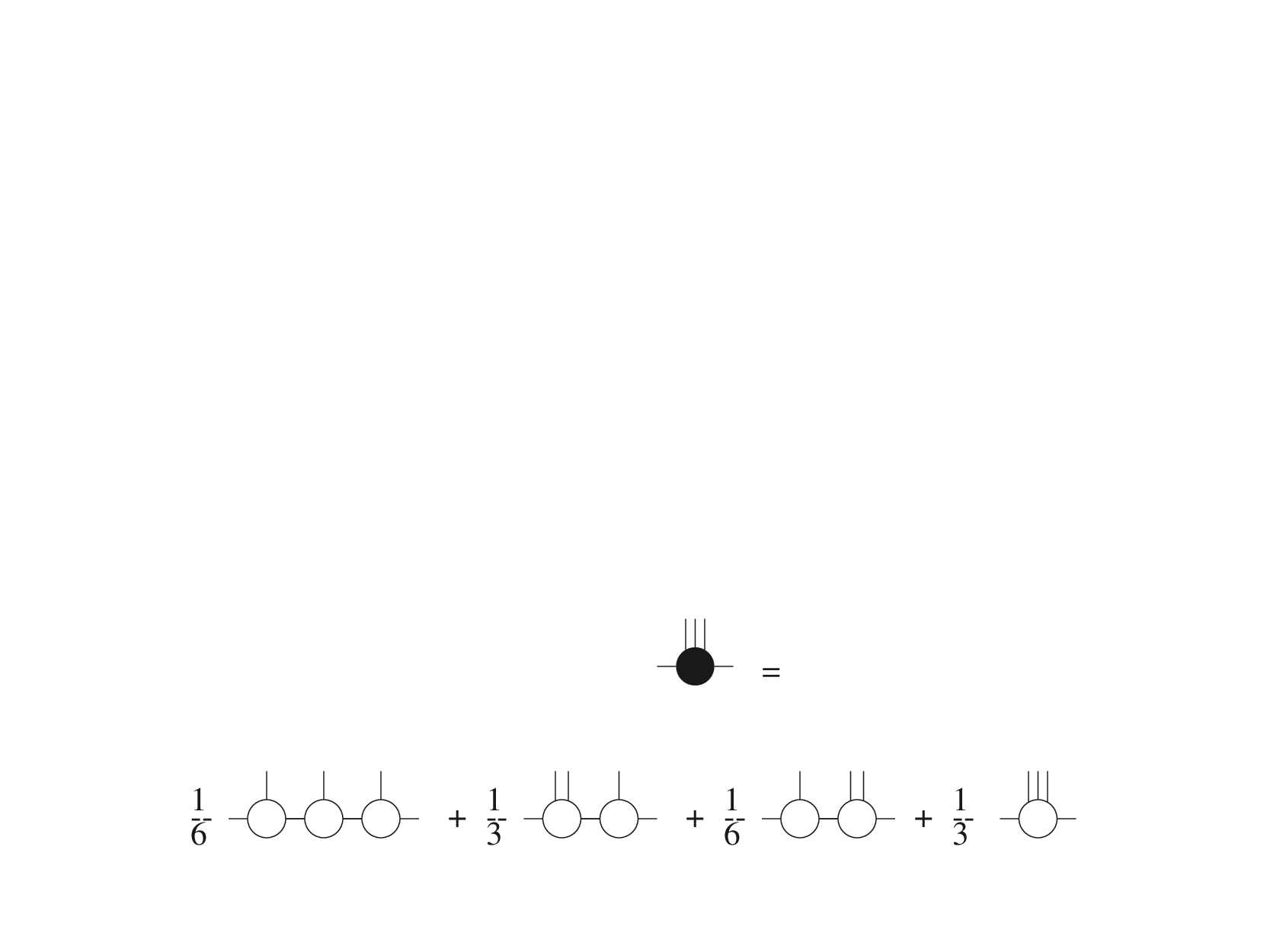}}
\nobreak
\vskip -0.5cm\nobreak
\caption{Graphical form for the decomposition of $U_3$ (filled circle)
in eq.~(\ref{ex2}).}
\end{figure}

\section{Exponential Formula for $U(T,T')$} 

The factorization character in (\ref{cor1}) 
and (\ref{ex2}) suggests that
it may be possible to sum up the power series $U_n$
to yield an explicit exponential function of the $C_n$'s.
This is indeed the case.

\subsection{Commutative $C_i$'s}

First assume all the $C_{m_i}$ in (\ref{cor1})
commute with one another. Then, as will be shown in Appendix C, 
\be
U[T,T']&=&1+\sum_{n=1}^\infty {U_n}\nn
&=&\prod_{j=1}^\infty \sum_{m=0}^\infty {1
\over j^{m}m!}C_j^{m}\nn
&=&\exp\[\sum_{j=1}^\infty{C_j\over j}\].\label{symexp}\ee

Explicit calculation of low-order terms
can also be obtained from (\ref{ex2}) for further verification. 
This yields
\be
U_0&=&1\nn
U_1&=&C_1\nn
U_2&=&{1\over 2}(C_1^2+C_2)\nn
U_3&=&{1\over 3!}C_1^3+{1\over 2}C_1C_2+{1\over 3}C_3\nn
U_4&=&{1\over 4!}C_1^4+{1\over 3}C_1C_3+{1\over 8}C_2^2+
{1\over 4}C_1^2C_2+{1\over 4}C_4\nn
U_5&=&{1\over 5!}C_1^5+{1\over 4}C_1C_4+{1\over 6}C_2C_3+
{1\over 6}C_1^2C_3\nn &+&{1\over 8}C_1C_2^2+{1\over 12}C_1^3C_2+{1\over 5}C_5
\nn
\sum_{i=0}^5U_n&=&\exp\[C_1+{1\over 2}C_2+
{1\over 3}C_3+{1\over 4}C_4
+{1\over 5}C_5\]
+R,\nn &&
\label{symexpex}\ee
where $R$ contains products of $C$'s whose subscript indices add up
to 6 or more.

\subsection{General Exponential Formula}

In general the $C_j$'s do not commute with one another so
the exponent in (\ref{symexp}) must be corrected by terms
involving commutators of the $C_j$'s. The exponent $K$
can be computed by taking the logarithm of $U(T,T')$:
\be
U(T,T')&=&1+\sum_{n=1}^\infty U_n=\exp[K]\equiv
\exp\[\sum_{i=1}^\infty K_i\],\nn
K&=&\ln\[1+\sum_{n=1}^\infty U_n\]=
\sum_{\ell=1}^\infty {(-)^{\ell-1}\over\ell}\[\sum_{n=1}^\infty
U_n\]^\ell\nn
&\equiv& \sum_{n=1}^\infty {C_n\over n}+\sum_m\eta(m)C_m,
\label{log}\ee
in which the last sum is taken over {\it all} $m=(m_1m_2\cdots m_k)$
for $k\ge 2$, and $C_m\equiv C_{m_1}C_{m_2}\cdots C_{m_k}$.
The resulting expression must be expressible
 as (multiple-)commutators of the 
$C$'s. Only commutators of $H(t)$, 
in the form of $C_m$ and their commutators, enter into $K$.
This is so because in the special case when $H(t)$ is a member of
a Lie algebra, $U(T,T')$ is a member of the corresponding Lie group
and so $K$ must also be a member of a Lie algebra.

By definition, $K_i$ contains $i$ factors of $H(t)$. 
Calculation for the first five is carried out in Appendix D. The result is
\be
K_1&=&\ \ C_1\nn
K_2&=&{1\over 2}C_2\nn
K_3&=&{1\over 3}C_3+{1\over 12}[C_2,C_1]\nn
K_4&=&{1\over 4}C_4+{1\over 12}[C_3,C_1]\nn
K_5&=&{1\over 5}C_5+{3\over 40}[C_4,C_1]+{1\over 60}[C_3,C_2]
+{1\over 360}[C_1,[C_1,C_3]]\nn
&+&{1\over 240}[C_2,[C_2,C_1]]
+{1\over 720}[C_1,[C_1,[C_1,C_2]]]\label{exp}\ee

$K_n$ consists of $C_n/n$, plus the compensating terms in the form of commutators
of the $C$'s. By counting powers of $H(t)$ it is clear that
 the subscripts of
these $C$'s must add up to $n$, but beyond that all independent
commutators and multiple commutators may appear.
For that reason it is rather difficult to obtain an explicit
formula valid for all $K_n$,
if for no other reason than the fact that new 
commutator structures appear in
every new $n$. It is however very easy to 
compute $K_n$ using (\ref{log}) in a computer.
This is actually how $K_5$ was obtained.

Nevertheless, when we stick to commutators of a definite structure,
their coefficients in $K$ can be computed as follows.

\subsection{Coefficient $\eta(m_1m_2\cdots m_k)$}

It is not difficult to compute $\eta(m_1\cdots
m_k)$ for small $k$ (but arbitrary $m_i$). The computation for
$k\le 4$ will be given below.
In order to avoid excessive subscripts we
shall use variables like $w,x,y,z$ to denote the positive integers
$m_i$. 

\subsubsection{$\eta(xy)$}

According to (\ref{log}), $C_xC_y$ can come from $U_{x+y}$, or $U_xU_y$.
The former corresponds to $\ell=1$, and the latter $\ell=2$. Using 
(\ref{ex2}), we obtain
\be
\eta(xy)={1\over (x+y)y}-{1\over 2xy}={y-x\over 2xy(x+y)}.\label{e2}\ee
The antisymmetry under $x,y$ exchange shows explicitly that it is
the commutator $\eta(xy)[C_x,C_y]$ that enters into $K$.

Using this formula, we can verify the coefficients of the 
single commutator terms appearing in (\ref{exp}): 
$\eta(2,1)=1/12,\ \eta(3,1)=1/12,\ \eta(4,1)=
3/40,\ \eta(3,2)=1/60$.

\subsubsection{$\eta(xyz)$}

According to (\ref{log}), $C_xC_yC_z$ can come from 
$U_{x+y+z}\
(\ell=1)$,  $U_xU_{y+z}$ and $U_{x+y}U_z\ (\ell=2)$, and $U_xU_yU_z\
(\ell=3)$. Using (\ref{ex2}), we get
\be
&&\eta(xyz)={1\over (x+y+z)(y+z)z}-{1\over 2}\biggl[{1\over x(y+z)z}\nn
&&+{1\over
(x+y)yz}\biggr]+{1\over 3}{1\over xyz}\nn
&&={y^2(x+z)+y(2x^2+2z^2-3xz)-xz(x+z)\over 6xyz(x+y)(y+z)(x+y+z)}.
\nn &&\label{e3}\ee
Because of the Jacobi identity there are only two independent double
commutators in $K$. They can be taken to be 
$\alpha[C_x,[C_y,C_z]]+\beta[C_y,[C_x,C_z]]$, if $y\not=z$.
We may take $\beta=0$ if $x=y$. In any case, we have
$\alpha=\eta(xyz)$. 

We can use this formula to verify the double commutator terms in
(\ref{exp}): $\eta(1,1,2)=1/360,\ \eta(2,2,1)=1/240$.

\subsubsection{$\eta(wxyz)$}

$C_wC_xC_yC_z$ may come from $U_{w+x+y+z}$, $U_{w+x+y}U_z$, 
{$U_{w+x}U_{y+z}$, $U_{w}U_{x+y+z}$, $U_{w+x}U_yU_z$, 
$U_wU_{x+y}U_z$,}\nobreak \break  $U_wU_xU_{y+z}$, and $U_wU_xU_yU_z$. Thus
\be
&&\eta(wxyz)=[(w+x+y+z)(x+y+z)(y+z)z]^{-1}\nn
&&-[
2(w+x+y)(x+y)yz]^{-1}
-[2(w+x)x(y+z)z]^{-1}\nn &&-[2w(x+y+z)(y+z)z]^{-1}
+[3(w+x)yz]^{-1}\nn
&&+[3w(x+y)yz]^{-1}
+[3wx(y+z)z]^{-1}
-[4wxyz]^{-1}.\nn &&\label{e4}\ee
It can be shown (Appendix E) that this is also the coefficient of
the triple commutator $[C_w,[C_x,[C_y,C_z]]]$.

This formula leads to $\eta(1,1,1,2)=1/720$, agreeing with the 
coefficient of the last term in (\ref{exp}).

\subsection{Gauge Invariance}

Under the `guage transformation' 
\be
\delta H(t)={d\Lambda(t)\over dt}+[\Lambda(t),H(t)]\equiv
\db H(t)+\da H(t),\label{gauge}\ee
where $\Lambda(t)$ is an arbitrary operator vanishing at
$t=T$ and $t=T'$, $U(T,T')$ is gauge invariant. This means that
$K=K(T,T')$ must also be gauge invariant. Since $\db$ decreases
the power of $H$ by one, and $\da$ leaves it unchanged, the gauge
invariance must be implemented order-by-order as
\be
\db U_{n+1}+\da U_n&=&0,\label{du}\\
\db K_{n+1}+\da K_n&=&0. \label{dk}\ee
The first equation is easy to verify but the second is not. See
Appendix F. The reason is that $U_n$ transforms simply under a gauge
change, as in (\ref{du}), but $C_n$ does not. It is this complexity
of gauge behaviour of $C_n$ that makes the verification of (\ref{dk})
complicated, but we will do it explicitly for the first few orders
in Appendix F. From this point of view, the reason why the dependence
of $K_n$ on $C_m$ is so complicated is simply because the complexity
is necessary to offset the complexity of the gauge behaviour
of $C_m$, so that the gauge change of $K_n$ becomes simple again,
as in (\ref{dk}).

\section{Formulas Resulting from Specific Choices of \boldmath{$H_i(t)$}}

It is well known that the path-ordered exponential $U[T,T']$
obeys the composition law 
\be U[T,T']=U[T,T'']U[T'',T']\label{comp}\ee
if $T\ge T''\ge T'$. We can combine this with a judicious choice
 of $H(t)$
to obtain many mathematical formulas. Three of them are presented below
for illustrative purposes. The first two are purely
mathematical results; the third is a nonabelian eikonal
formula useful for high-energy scattering amplitudes.

\subsection{Campbell-Baker-Hausdorff Formula}

Let $T=2,\ T''=1$, and $T'=0$. Let $H(t)=P$ for $2\ge t\ge 1$ and 
$H(t)=Q$ for $1>t\ge 0$, where $P$ and $Q$ are arbitrary matrices or
operators. Using (\ref{comp}) and (\ref{exp}), we obtain in this case (see Appendix G for details)
\be
C_m={1\over (m-1)!}(ad\ P)^{m-1}Q,\label{ad}\ee
where as usual $(ad\ P)V\equiv [P,V]$ for any operator
$V$. Substituting this into (\ref{exp}) we get
\be
\exp(P)\.\exp(Q)&=&\exp[K_1+K_2+K_3+K_4+K_5+\cdots]\nn
K_1&=&P+Q\nn
K_2&=&\h[P,Q]\nn
K_3&=&{1\over 12}[P,[P,Q]]+{1\over 12}[Q,[Q,P]]\nn
K_4&=&-{1\over 24}[P,[Q,[P,Q]]]\nn
K_5&=&-{1\over 720}[P,[P,[P,[P,Q]]]]\nn
&-&{1\over 720}[Q,[Q,[Q,[Q,P]]]]\nn
&+&{1\over 360}[P,[Q,[Q,[Q,P]]]]\nn
&+&{1\over 360}[Q,[P,[P,[P,Q]]]]\nn
&+&{1\over 120}[P,[Q,[P,[Q,P]]]]\nn &+&{1\over 120}[Q,[P,[Q,[P,Q]]]],
\label{BH}\ee
which is the Campbell-Baker-Hausdorff formula. The case
when $[P,Q]$ commutes with $P$ and $Q$ is well known. 
In that case all $K_n$ for $n\ge 3$ are zero.
Otherwise,
up to and including $K_4$ this formula 
can be found in eq.~(15), \S 6.4, Chapter II of Ref.~\cite{bourbaki}.

If necessary, we may also use the general knowledge for the coefficient
$\eta(m)$ obtained in Sec.~5.3 to deduce information
on higher-order terms.

\subsection{Translational Operator}
A simple and trivial example to illustrate the commutative formula
(\ref{symexp}) is obtained by choosing $T=3,\ T'=0$, 
$H(t)=a(d/dx)$ for $t\in[2,3]$, $H(t)=f(x)$ for $t\in[1,2]$, and
$H(t)=-a(d/dx)$ for $t\in[0,1]$. $f(x)$ is an arbitrary function of $x$
and $a$ is a constant. In this case
$C_n/n=(a^n/n!)(d^nf(x)/dx^n)$ (see Appendix H), 
so all commutators of $C_i$ vanish and (\ref{symexp}) is valid.
We get
\be
U[T,T']&=&\exp({ad/dx})\.\exp(f(x))\.\exp({-ad/dx})=\exp[K]\nn
&=&\exp\[\sum_{n=0}^\infty a^nf^{(n)}(x)/n!\]=\exp[f(x+a)].\label{taylor}\ee
This formula, expressing the fact that $\exp(ad/dx)$ is a translational
operator, is of course well-known. There is absolutely no need to derive
it with this heavy apparatus. We include it here just to illustrate how
different choices of $H(t)$ can lead to different formulas.

\subsection{Nonabelian Eikonal Formula}

A useful formula in high-energy scattering is obtained from (\ref{thm})
by choosing $H_i(t)=
\exp(ik_i\.pt)V_i\equiv\exp(i\o_it)$, 
where $V_i$ are time-independent operators, and $p,k_i$
are arbitrary four-momenta. We should also
choose $[T,T']$ to be $[\infty,-\infty]$ to correspond to `onshell amplitudes'
 and $[T,T']=[\infty, 0]$ to correspond to `offshell amplitudes'.

In that case, (\ref{thm}) reads
\be
U[123\cdots n]&=&a[123\cdots n]V_{1}V_{2}V_{3}\cdots V_{n}\nn
C[123\cdots n]&=&a[123\cdots n]\.\nn
&&\quad[V_{1},[V_{2},
[V_{3},[\cdots ,[V_{{n-1}},V_{n}]\cdots
]]]]\nn
a[123\cdots n]&=&i^{n-1}\(\prod_{j=1}^{n-1}{1\over\sum_{i=1}^j\o_i+i\e}\)
L_{[T,T']}\nn
L_{[\infty,-\infty]}&=&{2\pi\delta\(\sum_{i=1}^n\o_i\)}\nn
L_{[\infty,0]}&=&{1\over\sum_{i=1}^n\o_i+i\e}.\label{eik}\ee
Similar formulas can be obtained for $U[s]$ and $C[s]$ for any 
$s\in S_n$ by permutation. 
They are the tree amplitudes with vertex $V_i$ governing 
the emission of
$n$ bosons with momenta $k_i$ from a common source 
particle with momentum $p$.
The bosons are arranged along the tree
 in the order dictated by the permutation $s$. This formula is valid
 in the
high-energy approximation where $p^0\gg k_i^\mu$, whence spins can be ignored
and the product of propagators along the source line can be approximated by  
$a[s]$. $n!U_n$ is then the complete tree amplitude
obtained by summing the $n!$ tree diagrams. 
The decompostion theorem for $U_n$
in this case has been derived directly \cite{LL1}. Its onshell version
exhibits  spacetime and colour interference, and can be used to
prove the self-consistency of the baryonic
 scattering amplitudes in large-$N_c$
QCD \cite{LL2}. It also explains the emergence of reggeized gluons and the factorization of
high-energy near-forward scattering amplitudes into multiple-reggeon exchanges
\cite{FHL,FL}.

\section{Acknowlegements}

I am grateful to Herbert Fried, John Klauder, 
Cheuk-Yin Wong, and Tung-Mow Yan
for informative discussions and suggestions. 
This research is supported by the Natural Sciences and
Engineering Research Council of Canada. 

\newpage
\centerline{\Large{\bf APPENDICES}}

\appendix

\bigskip\bigskip

\section{Two Proofs of the General Decomposition Theorem}

On account of the importance of the decomposition theorem
(\ref{thm}) to the rest of this paper,
we shall provide two separate proofs for it.
Actually a third proof exists, along the lines
given in Ref.~\cite{LL1} for the
`multiple commutator formula' (which is simply the
`nonabelian eikonal formula' (\ref{eik})). However,
this requires the expansion of every operator $H_i(t)$
into a sum over a complete set of time-independent operators,
thus introducing lots of indices and even more complicated
notations. So we shall skip that third proof here.

The combinatorics needed for a general
proof are unfortunately rather involved,
though the basic idea
 in either case is really quite simple.
In order not to be bogged down with complicated
notations, we will start with the simplest case,
$n=2$, which already contains most of the basic ideas.
 
\subsection{\boldmath{$n=2$}}

\subsubsection{The Decomposition Theorem}

In that case, the decomposition 
theorem (\ref{thm}) reads
\be
2!U_2&=&U[12]+U[21]=C[1|2]+C[21]\nn
&=&C[1]C[2]+C[21]\label{thm21}
\\
U[12]&=&\int_{R[12]}dt_1dt_2H_1(t_1)H_2(t_2)\nn
&=&\int_{T'}^Tdt_1dt_2\theta(t_1-t_2)H_1(t_1)H_2(t_2)\nn &&
\label{thm22}\\
U[21]&=&\int_{R[21]}dt_1dt_2H_2(t_2)H_1(t_1)\nn
&=&\int_{T'}^Tdt_1dt_2\theta(t_2-t_1)H_2(t_2)H_1(t_1)\nn &&
\label{thm23}\\
C[1|2]&=&C[1]C[2]=\int_{T'}^Tdt_1dt_2H_1(t_1)H_2(t_2)
\label{thm24}\\
C[21]&=&\int_{R[21]}dt_1dt_2[H_2(t_2),H_1(t_1)]
\label{thm25}\ee

\subsubsection{First Proof for \boldmath{$n=2$}}

The first proof makes use of the simple observation that
the union of the two triangular integration regions, $R[12]=
\{T\ge t_1\ge t_2\ge T'\}$, and $R[21]=
\{T\ge t_2\ge t_1\ge T'\}$, is the square region
$\{T\ge t_1,t_2\ge T'\}$. Thus
\be
U[12]&+&U[21]\nn
&=&\int_{R[12]\cup R[21]}H_1(t_1)H_2(t_2)
+\int_{R[21]}[H_2(t_2),H_1(t_1)]\nonumber\\
&=&C[1]C[2]+C[21]\equiv C[1|2]+C[21].\label{pf21}\ee

\subsubsection{Second Proof for \boldmath{$n=2$}}

Eq.~(\ref{thm}) is clearly true when $T=T'$, for both
sides then vanish. To prove its general validity for any
$T'$, it is sufficient to show the 
$T'$-derivatives of both sides to be equal. For $n=2$, we have
to show that
\be
{d\over dT'}\{U[12]+U[21]\}={d\over dT'}\{C[1|2]+C[21]\}.
\label{dt20}\ee
This identity is true because
\be
{d\over dT'}U[12]&=&-\int_{T'}^TH_1(t)dtH_2(T')\label{dt21}\nn
{d\over dT'}U[21]&=&-\int_{T'}^TH_2(t)dtH_1(T')\label{dt22}\nn
{d\over dT}C[1|2]&=&-H_1(T')\int_{T'}^TH_2(t)dt
-\int_{T'}^TH_1(t)dt H_2(T')\label{dt23}\nn
{d\over dT'}C[21]&=&-[\int_{T'}^TH_2(t)dt,H_1(T')]\label{dt24}.\ee

We shall now proceed to the general proofs.

\subsection{First Proof for Arbitrary \boldmath{$n$}}

In the case of $n=2$, the `first proof' involves
two crucial steps. Step one is to recognize that
the union of the two triangular regions, $R[12]$
and $R[21]$, is a square, thus allowing the first
term on the right-hand-side of (\ref{pf21}) to be
factorized. The generalization of this to an arbitrary
$n$ is the subject of discussions immediately follows.
The second crucial step is to introduce commutators 
to rewrite $H_2(t_2)H_1(t_1)$
as a sum of two terms: $H_1(t_1)H_2(t_2)+
[H_2(t_2),H_1(t_1)]$. The generalization of this step
to  arbitrary $n$ will then be discussed. In a final
subsection, these two steps will be assembled 
to complete the proof of the
general decomposition theorem.

\subsubsection{Sums and Cartesian Products of
Hyper-triangular Regions}

Let $[s]$ be a sequence of $n$ natural numbers, 
not necessarily consecutive, and $[s]=
[\tilde s_1\tilde s_2\cdots \tilde s_p]$ be a partition
of this sequence into $p$ subsequences. 
For example, $[s]=[3254167]$ could be partitioned into
three subsequences $[\tilde s_1]=[32],\ [\tilde s_2]=[5416]$,
and $[\tilde s_3]=[7]$. We define the {\it set} 
$\{\tilde s_1;\tilde s_2;\cdots;\tilde s_p\}$ to consist
of all sequences of these $n$ natural numbers,
obtained by merging and interleaving
the numbers in $\tilde s_1,\tilde s_2,
\cdots,\tilde s_p$ in all possible ways, 
subject to the condition that the
original orderings of numbers within each $\tilde s_i$ be
kept fixed.
This set then contains $n!/(k_1!k_2!\cdots k_p!)$ sequences,
where $k_i$ is the number of numbers in the subsequence
$\tilde s_i$, and $\sum_{i=1}^pk_i=n$. In the explicit
examples above, the set $\{32;5416;7\}$ will consist
of $7!/(2!4!1!)=105$ sequences, which among others include
$[3254167],\ [3524176]$, and $[7534162]$, 
but {\it not} $[2354167]$ and $[6541732]$ because the
original orderings are not kept in these last two cases: 
the numbers 2 and 3 are
 reversed in the first instance and the number 6
appears before the numbers 5,4,1 in the second.

Given a permutation $s\in S_n$, recall that the hyper-triangular
integration region $R[s]$ is defined to be $\{T\ge t_{s_1}
\ge t_{s_2}\ge\cdots\ge t_{s_n}\ge T'\}$. 
We shall now define the integration region $R\{\tilde s_1;
\tilde s_2;\cdots;\tilde s_p\}$ to be the union of all
the regions $R[t]$ with $[t]\in\{\tilde s_1;\tilde 
s_2;\cdots;\tilde s_p\}$. A moment's thought reveals
that this is nothing but the Cartesian product of the
individual hyper-triangular regions $R[\tilde s_1],
R[\tilde s_2],\cdots, R[\tilde s_p]$. This leads to
a {\it factorization theorem} for the following sums of 
$U[t]$:
\be
\sum_{[t]\in\{\tilde s_1;
\tilde s_2;\cdots;\tilde s_p\}}U[t]&=&
\int_{R\{\tilde s_1;
\tilde s_2;\cdots;\tilde s_p\}}dt_1dt_2\cdots dt_n\nn
&&H_{s_1}(t_{s_1})
H_{s_2}(t_{s_2})\cdots H_{s_n}(t_{s_n})\nn
&=&U[\tilde s_1]U[\tilde s_2]\cdots U[\tilde s_p].
\label{fact}\ee
A factorization theorem similar to this will be used for
the proof of the decomposition theorem.

\subsubsection{Canonical Ordering}
From now on we shall denote $H_{s_i}(t_{s_i})$ simply
as $H_{s_i}$, and $H_{s_1}H_{s_2}\cdots H_{s_n}$
as $H[s_1|s_2|\cdots|s_n]\equiv H_c[s]$. 
The subscript `$c$' of $H$ serves to 
indicate that a vertical
bar is to be put after each number in the sequence
$[s]$ to indicate that products of operators shuold be taken. 

Similar symbol without
the verti~cal bar will  denote {\it nested multiple commutators}. Thus~
$H[1|2|3|4]=H_1H_2H_3H_4=H_c[1234]$, but 
$H[12|34]=[H_1,H_2]\.$ $[H_3,H_4]$
and $H[1234]=[H_1,[H_2,[H_3,H_4]]]$, etc. 

Suppose $[s]=[s_1s_2\cdots s_n]$ is a sequence of $n$
numbers, not necessarily consecutive. Then $[s]_c$
will be used to denote the same sequence, but with
{\it vertical bars inserted after $s_i$ iff $s_i<s_j$ for all
$i<j$}. For example, if $[s]=[7125]$, then $[s]_c=[71|2|5]$,
and if $[s]=[21345]$, then $[s]_c=[21|3|4|5]$.
An operator of the form $H[s]_c$ will be called a {\it canonical operator}. Please note the difference between
$H_c[s]$ and $H[s]_c$. The former is generally not a
canonical operator, but the latter is, by definition.

Let $s\in S_n$. From (\ref{Us}), 
the integrand in $U[s]$ is given by
$H_c[s]$, 
which as mentioned above is generally {\it not}
a canonical operator. By introducing commutators,
it is possible to {\it canonically order} this operator,
meaning that it can be written
as a linear combination of canonical
operators. I shall illustrate how this is done by
looking at a simple example, $H[3|2|1]=H_3H_2H_1=H_c[321]$:
\be
&&H[3|2|1]=H_3H_2H_1\nn
&&=H_3[H_2,H_1]+[H_3,H_1]H_2
+H_1H_3H_2\nn
&&=[H_2,H_1]H_3+[H_3,[H_2,H_1]]+[H_3,H_1]H_2+H_1H_3H_2\nn
&&=H[21|3]+H[321]+H[31|2]+H[1|3|2]\label{can31}\\
&&=H[21|3]+H[321]+H[31|2]+H[1|2|3]+H[1|32].\nn &&\label{can32}\ee
Note that the first three terms of (\ref{can31})
are canonical operators but not the fourth.
This is then fixed in (\ref{can32}) so that all its 
five operators are now canonical.

With this example in mind, we can now discuss how to
canonically order an arbitrary operator of the form $H_c[s]$,
where $[s]$ is a sequence of numbers, not necessarily
consecutive.

Locate the smallest number $s_0$ in the sequence.
By introducing commutators and commutators of commutators,
we shall try to move $H_{s_0}$, and commutators 
involving it, 
 to the extreme left. This results in a sum of a number
of terms which can be described as follows.

Suppose $s_0$ is situated somewhere in the middle of $[s]$
so that $[s]=[\tilde s_1s_0\tilde s_2]$,
with a subsequence $[\tilde s_1]$ before $s_0$ and a 
subsequence $[\tilde s_2]$ after. Let
$[\tilde s'_1]$ and $[\tilde s''_1]$ be two complementary
subsequences
of numbers in $[\tilde s_1]$ in the sense that all numbers
in $[\tilde s_1]$ appear either in $[\tilde s'_1]$
or $[\tilde s''_1]$. For example, if $[\tilde s_1]
=[7352]$, then $[\tilde s'_1],\ [\tilde s''_1]$ may be
$[3],\ [752]$, or $[75],\ [32]$, etc., but not
$[7], [52]$ (3 is left out), nor $[37],\ [52]$ (order of 7,3
are reversed). If $[\tilde s_1]$ has $q$ numbers, then
there are altogether $2^q$ pairs of complementary subsequences
like that. 

Now we are ready to describe the terms obtained by
the move of $H_{s_0}$ and its commutators to the extreme left
in order to achieve canonical ordering. They involve all terms
of the form $H[\tilde s'_1s_0]H_c[t]$, where
$[t]=[\tilde s''_1\tilde s_2]$ is the sequence obtained
by prepending
  the subsequence $[\tilde s''_1]$ to the subsequence
$[\tilde s_2]$. Since $s_0$ is the smallest number
in $[s]$, the operator $H[\tilde s'_1s_0]$ is canonical,
though the remaining factor $H_c[t]$ may not be.
If not, we will then repeat the same procedure, 
find the smallest
number $t_0$ of $[t]$, and move $H_{t_0}$ and its
commutators to the extreme left of $H_c[t]$ to render
 sections of $H_c[t]$ canonical. Continuing thus,
eventually $H_c[s]$ can be rendered into sums of 
canonical operators. 

We can describe this result
 in another  way, more useful for our subsequent proof.
Given a canonical operator $H[u]_c$, with $[u]\in S_n$,
this result states that it will appear in the canonical
ordering of  $H_c[s]$
iff $[s]\in\{u\}$, where $\{u\}$ is the {\it set} of
permutations obtained by changing the vertical bars
in $[u]_c$ into semicolons `;', and the square brackets
$[\cdots]_c$ into curly brackets $\{\cdots\}$.
For example, if $[u]_c=[21|3]$, then $\{u\}=\{21;3\}
=\{[321],[231],[213]\}$, and the operators $H_c[s]$
containing the canonical operator $H[21|3]$ are
$H_c[321]=H[3|2|1]$, $H_c[231]=H[2|3|1]$, and
$H_c[213]=H[2|1|3]$. 

\subsubsection{Proof}

We can now assemble these two ingredients into a proof
of the general decomposition theorem (\ref{thm}). According
to the rule for canonical ordering, the integrand
$H[u]_c$ will be contained in $U[s]$ iff
$[s]\in\{u\}\equiv\{\tilde u_1;\tilde u_2;\cdots;
\tilde u_q\}$. The integration region for the integrand
$H[u]_c$ is therefore 
$\bigcup_{[s]\in\{u\}}R[s]=R\{u\}$. The resulting
integral is therefore
\be
&&\int_{R\{u\}}dt_1\cdots dt_n H[\tilde u_1|\tilde u_2|
\cdots|\tilde u_q]\nn
&=&C[\tilde u_1]C[\tilde u_2]\cdots C[\tilde u_q]
=C[\tilde u_1|\tilde u_2|\cdots|\tilde u_q].\label{pfn1}
\ee
The last step  is similar to the one used to
obtain the factorization theorem (\ref{fact}).
Summing over all possible $[u]\in S_n$ is equivalent to
summing over all possible $s\in S_n$, by doing
so  we will get
from (\ref{pfn1}) the general decomposition theorem (\ref{thm}).

\subsection{Second Proof for Arbitrary $n$}

The decomposition theorem is  trivially true when
$T=T'$, thus its general validity would follow if the
 $T'$-derivatives of (\ref{thm}) is obeyed:
\be
{d\over dT'}\sum_{[u]\in S_n}U[u]={d\over dT'}
\sum_{[u]\in S_n}C[u]_c.\label{thmn2}\ee
Our second proof consists of proving this equation, with the 
help of the identities
\be
{d\over dT'}U[u]&=&-U[u']H_r(T'),\label{pfn21}\\
{d\over dT'}C[\tilde u_j]&=&-[C[\tilde u'_j],H_s(T')],\label{pfn22}
\ee
where $r$ is the last element in $[u]\in S_n$ and 
$s$ is the
last element of $[\tilde u_j]$. In other words,
$[u]=[u'r]$ and
$[\tilde u_j]=[\tilde u'_js]$.  $[\tilde u_j]$ is one of
the sequences between vertical bars in $[u]_c$, and by
definition $C[u]_c$ is given by a product of $C[\tilde u_j]$
over different $j$.

We have to be careful with 
(\ref{pfn21}) and 
(\ref{pfn22}) when $[u]$ or $[\tilde u_j]$
contains only one number. In that case 
$[u']$ and $[\tilde u'_j]$ are void and $U[u']=1$. 
Moreover,
\be
{d\over dT'}C[s]=-H_s(T').\label{pfn23}\ee

Let $S_{n-1}[m]$ be the permutation group of the first $n$ natural
numbers with the number $m$ removed. Using (\ref{pfn21}), the
left-hand-side of (\ref{thmn2}) becomes 
\be
{d\over dT'}\sum_{[u]\in S_n}U[u]&=&-\sum_{m=1}^n 
\sum_{[u']\in S_{n-1}[m]}U[u']H_m(T').\label{pfn24}\ee
In order for (\ref{thmn2}) to be true, the right-hand-side must
also be given by (\ref{pfn24}).

Now in (\ref{pfn24}) 
the operators $H_m(T')$ always appear at the extreme right.
On the other hand, if we apply (\ref{pfn22},\ref{pfn23}) to compute 
the right-hand-side of (\ref{thmn2}), the operators $H_m(T')$
may also appear at other positions.  
In fact, when $H_m(T')$ appears
$m$ is always
the last element of $[\tilde u_j]=[\tilde u'_jm]$ for some $j$. 
To determine for what $[u]_c$ this occurs we consider an arbitrary
element $[u']\in S_{n-1}[m]$. Suppose $[u']_c=[\tilde u'_1|
\tilde u'_2|\cdots|\tilde u'_k]$. We will now discuss where in $[u']_c$
we may insert the number $m$ 
to obtain a legitimate $[u]_c$ for some $[u]\in S_n$ in which $m$
appears just before a vertical bar, or the right-hand square bracket $]$.
If $m$ is to be inserted at the end of a subsequence $\tilde u'_j$,
then $m$ must be smaller than all the numbers in $\tilde u'_j$.
Since the last element of $\tilde u'_j$ increases with $j$, if
$m$ may appear at the end of $\tilde u'_j$ it may also appear at the
end of $\tilde u'_\ell$ for all $\ell>j$. On the other hand, there
may be a smallest $j=j_0$. In that case, it is not allowed to
insert $m$ at the end of $[\tilde u'_j]$ for $j<j_0$, but the insertion
$[\tilde u'_1|\cdots|m|\tilde u'_{j_0}|\cdots |\tilde u_k]$
is allowed.

So consider a fixed $[u']\in S_{n-1}[m]$ and all the resulting 
$[u]\in S_n$ with these insertions of the number $m$. Using (\ref{pfn22})
and (\ref{pfn23}), it is now easy to see that $dC[u]_c/dT'$,
summed over all such $[u]$'s, will give $-C[u']_cH_m(T']$.
Summing over all $[u']$ and all
$m$, we therefore regain the right-hand-side of
(\ref{pfn24}), and hence proving (\ref{thmn2}), provided
\be
\sum_{[u']\in S_{n-1}[m]}U[u']=\sum_{[u']\in 
S_{n-1}[m]}C[u']_c.\label{pfn25}\ee
This last identity follows from the induction hypothesis (on $n$), which
we may invoke, because (\ref{thm}) is true for $n=2$. This completes
the second proof of the decomposition theorem.

\section{Special Decomposition Theorem}

When the operator functions $H_i(t)$ are all identical, the decomposition
theorem reduces to eq.~(\ref{cor1}), as we shall show by deriving
the expression for $\xi(m)$. This is equal to the number of $[s]_c
=[\tilde s_1|\tilde s_2|\cdots|\tilde s_k]$ with $[\tilde s_j]$
containing $m_j$ numbers ($\sum_jm_j=n$), divided by $n!$.

We remind the readers how to construct
$[s]_c$ for every $s\in S_n$.
A vertical bar is placed
behind a number $s_i$ iff $s_i<s_j$ for all $i<j$. 

Thus the last element of $[\tilde s_1]$ is always the smallest 
of the $n$ numbers, {\it i.e.,} the number `1'. There are therefore
$[(n-1)!/(m_1-1)!(n-m_1)!](m_1-1)! = n!/(n-m_1)!$ ways of constructing
$[\tilde s_1]$.

Similarly, the
last element of $[\tilde s_2]$ is the smallest number in the
residue sequence $[s]/[\tilde s_1]=[\tilde s_2\t s_3\cdots \t s_k]$. 
The remaining $m_2-1$ numbers in $\t s_2$ can be chosen 
arbitrarily from the $n-1-m_1$ elements in the residue sequence,
so there are $(n-1-m_1)!/(n-m_1-m_2)!$ ways of doing so. 
Continuing thus, the
total number of ways is simply the product of these numbers.
After dividing by $n!$, we get
\be
&&\xi(m_1m_2\cdots m_k)\nn
&&=[(n-m_1)(n-m_1-m_2)\cdots
(n-\sum_{i=1}^{k-1}m_i)]^{-1},
\ee
which is the same as the formula given in (\ref{cor1}).

\section{Exponential Formula for Commuting $C_i$'s}

To obtain (\ref{symexp}) when all the $C_i$'s commute, we need to
know the number of $[s]_c$ whose vertical bars separate the sequence
into $m_j$ subsequences of length $j$ ($j=1,2,3,\cdots$).

The number of ways that $n$ numbers can be divided into $m_j$
groups of $j$ numbers is $n!/\prod_j(m_j!j!^{m_j})$. From any
one of these groupings we can produce $\prod_j(m_j-1)!$ viable
$[s]_c$'s in the following way. Recall that the vertical bars of
$[s]_c$ are put behind a number $s_i$ iff $s_i<s_k$ for all $i<k$.
Stated differently, this means that (i) the last number in each group
of numbers separated by the vertical bars is always the smallest number
of that group, and that (ii) the groups must be arranged in such a
way that their smallest numbers increase from left to right. 
For a given grouping of numbers, there are $\prod_j(j-1)!^{m_j}$ ways
of arranging the orderings within each group to satisfy (i).
Once this is done, we can order the different groups in a unique
way so that (ii) is satisfied. Hence the number of $[s]_c$
is given by $[n!/\prod_j(m_j!j!^{m_j})][\prod_j(j-1)!^{m_j}]
=n!/(\prod_jm_j!j^{m_j})$. From this the rest of (\ref{symexp})
follows easily.

\section{General Exponential Formula}

(\ref{exp}) can be computed from (\ref{log}) and (\ref{cor1}) as follows.
If $P(U_n)$ is a polynomial of $U_n$, then $\[P(U_n)\]_m$ is obtained
from $P(U_n)$ by discarding all products of $U_i$'s
whose `degree' (sum of indices
$i$) is not equal to
$m$.  With this convention we have
$K_n=\sum_{\ell=1}^n(-)^{\ell-1}[U^\ell]_n$, so
\be
K_1&=&U_1=C_1\nn
K_2&=&=U_2-{1\over 2}U_1^2
={1\over 2}C_2\nn
K_3&=&
U_3-{1\over 2}(U_1U_2+U_2U_1)
+{1\over 3}U_1^3\nn
&=&{1\over 3}C_3+{1\over 12}[C_2,C_1]\nn
K_4&=&U_4-{1\over 2}\[U_1U_3+U_3U_1+U_2^2\]
+{1\over 3}\bigl[U_1^2U_2\nn
&+&U_1U_2U_1
+U_2U_1^2\bigr]-{1\over 4}U_1^4\nn
K_5&=&U_5-{1\over 2}\[U_1U_4+U_4U_1+U_2U_3+U_3U_2\]
+{1\over 3}[U_1^2U_3\nn
&+&U_1U_3U_1+U_3U_1^2
+U_2^2U_1+U_2U_1U_2+U_1U_2^2]\nn
&-&{1\over 4}\bigl[U_1^3U_2\nn
&+&U_1^2U_2U_1+U_1U_2U_1^2+U_2U_1^3\bigr]
+{1\over 5}U_1^5.
\ee
Substituting in the expressions given in (\ref{ex2}) for $U_i$
into the last two equation,
we obtain the
expression for $K_4$ and $K_5$ given in (\ref{exp}).

\section{Nested Multiple Commutators}

Given $n$ operators $C_{m_i}\equiv B_i$, we denoted their nested
multiple commutators by
\be
B[s]&\equiv& B[s_1s_2\cdots s_n]\nn
&\equiv&[B_{s_1},[B_{s_2},
[\cdots,[B_{s_{n-1}},B_{s_n}]\cdots]]],\label{nested}\ee
where $s\in S_n$. We want to show that the $(n-1)!$ nested operators
$B[s'1]$, with $s'\in S_{n-1}[1]$ (this is defined just below 
eq.~(\ref{pfn23})) forms a linear basis for all the $n$-tuple multiple
commutators. In other words, if $V_n$ is the vector space spanned
by these $(n-1)!$ nested commutators, then $V_n$ contains all
$n$-tuple multiple commutators.

This means that (i) multiple commutators not of the nested form can be
written as linear combinations of the multiple commutators of the nested
form, and (ii) nested multiple commutators with $B_1$ not at the
last place can be written as linear combinations of those with $B_1$
at the last place.

We shall use $M_n$ to denote the vector space generated by
all $n$-tuple multiple commutators. We shall now proceed to show
that $M_n=V_n$.

To denote a multiple commutator not of the nested type, we use
parentheses to single out those factors within  it that are
nested.  For example, 
\be
B[((123)4(567))8]&=&[B[(123)4(567)],B_8]\nn
B[(123)4(567)]&=&B[(123)4567]\nn
&=&[B[(123)],[B_4,[B_5,[B_6,B_7]]]]\nn
B[(123)]&=&B[123]=[B_1,[B_2,B_3]].\ee
Note that if $[s]$ is any sequence of numbers, then $B[\cdots(s)]
=B[\cdots s]$. In other words, the pair of parentheses 
located at the end can be dropped.

To prove (i), we must show that all the parentheses can be removed.
This can be accomplished via the Jacobi
identity
\be[A_1,[A_2,A_3]]=[A_3,[A_2,A_1]]-[A_2,[A_3,A_1]],\label{j3}\ee
with $A_1$ being the operator contained in
 the leftmost pair of parentheses.
The identity will be used to
move $A_1$ to the right. If it reaches the end then
the parenthses can be dropped.
Repeating this procedure eventually we can gradually
move all the parentheses
to the end and have them all dropped. This shows that $M_n=V_n$.

We shall now prove (ii) by induction.

For $n=2$ this is obvious because 
\be [B_2,B_1]=-[B_2,B_1].\label{j2}\ee

For $n=3$ this follows from (\ref{j2}) and the Jacobi identity
(\ref{j3}). We merely have to take $B_i=A_i$.

Assuming this can be done for $n$ up to $n=m-1$, we must now show
it to be true for $n=m$. 

Unless $B_1$ is at the beginning position there is nothing to prove,
for otherwise $B_1$ and operators to its right are located in
an $M_n=V_n$ for $n\le m-1$, so $B_1$ can be moved to the end.

If $B_1$ is located at the first postion, so that the nested
multiple commutator is of the form $B[1s]$ for some sequence $[s]$
with $n-1$ numbers, then using (\ref{j3}) with $B_1=A_1$ we can 
move $B_1$ to the right. Again it and the operators to its right now
belongs to $M_n$ with $n\le m-1$, so by the induction hypothesis
$B_1$ can be moved to the end. This completes the proof of (ii).

\mbox{In particular, $V_4$ is  spanned by the 6 nested commutators
$B[4321],\ B[4231],\ B[3421],$} $B[3241],\ B[2431],$ and
$B[2341]$. Only the first nested commutator contains the term
$B_4B_3B_2B_1$, so the coefficient of this nested commutator in $K$
is identical to the coefficient of this term in $K$, a fact that
has been used in Sec.~5.3.3 to compute the coefficient of a 
nested commutator.

This result can also be used to understand why only the particular
nested commutators of $H_i$ defined in $C_m$ occurs in $K$.

\section{Gauge Invariance}

We want to study in this Appendix the behaviour of $U_n$, $C_n$,
and $K_n$
under the `gauge transformation'
\be
\delta H(t)={d\Lambda(t)\over dt}+[\Lambda(t),H(t)]\equiv
\db H(t)+\da H(t),\label{gaugea}\ee
where $\L(t)$ is an arbitrary function of $t$ 
vanishing at the boundaries: $\Lambda(T)=\Lambda(T')=0$. 

Recall that $U_n=U[123\cdots n]$ and $C_n=C[123\cdots n]$,
that time is ordered according to the sequence of indices
 in the square bracket,
and that an operator $H(t_i)$ is present inside the integral at the
position of the index $i$. The only difference between $U[\cdots]$
and $C[\cdots]$ is that this operator appears as a factor
in a straight product in the former, and it appears inside a
nested multiple commutator in the latter.

We shall now introduce an index with a prime to denote the operator
$\Lambda$. At the position where $i'$ appears in the sequence
will be the operator $\Lambda(t_i)$. If the primed index is inside
$U[\cdots]$, then $\Lambda$ is a member of a straight product as before.
If it appears inside $C[\cdots]$, then it is a part of the nested
commutators.

We will also introduce parenthesis $(i'i)$ to denote commutators.
Thus at the position this appears should stand the operator
$[\Lambda(t_i),H(t_i)]$. 

With these notations we are now ready to discuss the gauge properties
of the various quantities. 

For $U_n=U[123\cdots n]$ it is simple:
\be
\da U_{n+1}&=&\sum_{j=1}^n\{-U[12\cdots j'j\cdots n]+
U[12\cdots jj'\cdots n]\}\nn
&=&-\sum_{j=1}^n
U[12\cdots (j'j)\cdots n],\label{du1}\\
\db U_n&=&\sum_{j=1}^nU[12\cdots (j'j)\cdots n].\label{du2}\ee

 Clearly (\ref{du}) is satisfied and $U$ is gauge invariant.

Now for $C_n$:
\be
\da C_{n+1}=\sum_{j=1}^n\{&-&C[12\cdots j'j\cdots n]\nn
&+&
C[12\cdots jj'\cdots n]\}\label{dc1}\\
\db C_n&=&\sum_{j=1}^nC[12\cdots (j'j)\cdots n],\label{dC2}\ee
so they look deceptively similar to the equations for $U_n$. 
However, since the operators in $C[\cdots]$ appear in {\it nested}
commutators, the two terms in (\ref{dc1}) can be combined into
$-C[12\cdots (j'j)\cdots n]$ only with the help of the Jacobi identity,
and this we can do only for $j\le n-1$. Jacobi identity involves
double commutators and those are absent at $j=n$.
Instead, we can use the antisymmetry of a single commutator to add up
these two terms. Hence
\be
\da C_{n+1}&=&-\sum_{j=1}^{n-1}C[12\cdots (j'j)\cdots n]
-2C[12\cdots (n'n)].\nn &&\label{dc3}\ee
The parenthesis in the last term may be dropped; we kept it there
just for the uniformity of notation with the other terms. The main
difference between this and (\ref{du1}) is that this last term
now has a factor of 2, which makes $\db C_{n+1}+\da C_n\not=0$.
It is this `slight' difference that eventually makes $K$ very
complicated just in order to keep its gauge invariance according
to (\ref{dk})!

We shall now use (\ref{dc2}) and (\ref{dc3}) to verify (\ref{dk})
for the first few orders. Since $K_n$ contains the term $C_n/n$,
it is useful to compute
\be
&&{1\over n+1}\db C_{n+1}+{1\over n}\da C_n\nn
&&=
{1\over n(n+1)}\{\sum_{j=1}^{n-1}C[12\cdots (j'j)\cdots n]\nn
&&\qquad -(n-1)C[12\cdots (n'n)]\}.\label{dc4}\ee
Referring back to (\ref{exp}), commutator terms are not present
at $K_n$ for $n=1,2$, which means that gauge invariance of $K$
demands (\ref{dc4}) to be zero for $n=1$, which it is.
For $n=2$, 
\be
&&\db K_3+\da K_2={1\over 3}\db C_3+
{1\over 2}\da C_2+{1\over 12}\db\([C_2,C_1]\)\nn
&=&{1\over 6}\(C[(1'1)2]-C[1(2'2)]-[C[(1'1)],C_1]\).\label{dc5}\ee
The last expression is arrived at by using (\ref{dc4}),\  
$\db C_1=0$, and $\db C_2=-2C[(1'1)]$. Now use the observation
just before eq.~(\ref{pf21}) to obtain the identity
\be
[C_1,C[(1'1)]]=C[1(2'2)]-C[(1'1)2].\ee
Substituting this back into (\ref{dc5}) we conclude that
$\db K_3+\da K_2=0$.

I have verified (\ref{dk}) to one higher order. The verification
becomes increasingly more difficult at larger $n$.

\section{Campbell-Baker-Hausdorff Formula}

Consider a special case in which $T=2,\ T'=0$, 
$H(t)=P$ for $2\ge t\ge 1$,  and $H(t)=Q$ for 
$1\ge t\ge 0$.
The operators $P$ and $Q$ are completely arbitrary.

We consider the ramification of
(\ref{exp}) in this special case. Using (\ref{comp}) with $T''=1$, 
the left-hand-side of (\ref{exp}) is just $\exp(P)\exp(Q)$.
To compute the right-hand-side we must first compute the $C_m$'s in
this special situation using (\ref{Cs}).

Clearly $C_1=P+Q$. For $C_m$ with $m\ge 2$, in order for the commutators
inside (\ref{Cs}) not to vanish we must have $t_m$ between 0 and 1,
and all the other $t_i\ (1\le i\le m-1)$ between 1 and 2. Thus
$C_m=(ad\ P)^{m-1}Q/(m-1)!$, where as usual $(ad\ P)V=[P,V]$ for
any operator $V$. Substituting this back into (\ref{exp})
we get the Campbell-Baker-Hausdorff formula shown in (\ref{BH}).

\section{Translational Operator}

Consider now the special case when $T=3,\ T'=0$, $H(t)=ad/dx$
for $t$ between 3 and 2, $H(t)=f(x)$ for $t$ between 2 and 1, and 
$H(t)=-ad/dx$ for $t$ between 1 and 0. Here $f(x)$ is an arbitrary
function and $a$ is a constant.

By using (\ref{comp}) the left-hand-side of (\ref{exp}) becomes
$\exp(ad/dx)\cdot\exp(f(x))\cdot$\break$\exp(-ad/dx)$. To compute the right-hand-side
we need to compute $C_m$ using (\ref{Cs}). It is clear that $C_1=f(x)$.
To compute $C_{m+1}$ for $m\ge 1$, note the following. 
In order for the commutators
in (\ref{Cs}) not to vanish, $t_{m+1}$ must either be between 0 and 1, 
or between
1 and 2. We shall denote the former contribution by $C_{m+1}'$, and
the latter by $C_{m+1}''$.

To compute $C_{m+1}'$, we must have $t_{m}$ to be between 1 and 2,
and $t_i$ for $i<m$ to be between 2 and 3. Hence 
$C_{m+1}'=a^m(d^mf(x)/dx^m)/(m-1)!$. For $C_{m+1}''$,
we must have all $t_i$ for $i\le m$ to be between 2 and 3, hence
$C_{m+1}''=a^m(d^mf(x)/dx^m)/m!$. Adding up the two, we get
$C_{m+1}=C'_{m+1}+C''_{m+1}=(m+1)a^m(d^mf(x)/dx^m)/m!$. Hence all
the $C_i$'s commute with one another, so (\ref{symexp}) can be 
used instead. (\ref{taylor}) then follows immediately.


\begin{thebibliography}{9}
\bibitem[*]{email} Email: Lam@physics.mcgill.ca
\bibitem{dragt} A.J. Dragt and J.M. Finn, {\it J. Math. Phys.}
{\bf 17} (1976) 2215; A.J. Dragt and E. Forest, {\it ibid},
{\bf 24} (1983) 2734; L.M. Healy, A.J. Dragt, and I.V. Gjala,
{\it ibid} (1992) 1948.
\bibitem{EIK} R. Torgerson, {\it Phys. Rev.} {\bf 143} (1966) 1194;
H. Cheng and T.T. Wu, {\it Phys. Rev.} {\bf 182} (1969)
1868, 1899; M. Levy and J. Sucher, {\it Phys. Rev.} {\bf 186} (1969) 1656.
\bibitem{geom} For a review, see H. Cheng and T.T. Wu, {\it
`Expanding Protons: Scattering at High Energies'}, (M.I.T. Press, 1987);
R.J. Glauber, in {\it `Lectures in Theoretical Physics'},
ed. W.E. Brittin and L.G. Dunham (Interscience, New York, 1959, vol. 1).
\bibitem{IR} D. Yennie, S. Frautschi and H. Suura, {\it Ann. Phys.}
{\bf 13} (1961) 379; S. Weinberg, {\it Phys. Rev.} {\bf 140} (1965) B516;
G. Grammer, Jr. and D.R. Yennie, {\it Phys. Rev. D} {\bf
8} (1973) 4332.
\bibitem{LL1}
C.S. Lam and K.F. Liu, {\it Nucl. Phys.} {\bf B483} (1997) 514.
\bibitem{LL2} C.S. Lam and K.F. Liu, {\it Phys. Rev. Lett.}
{\bf 79} (1997) 597. 
\bibitem{FHL}Y.J. Feng, O. Hamidi-Ravari, and C.S. Lam,
{\it Phys. Rev. D} {\bf 54} (1996) 3114.
\bibitem{FL} Y.J. Feng and C.S. Lam, {\it Phys. Rev. D}
{\bf 55} (1997) 4016.
\bibitem{LPM} L.D. Landau and I.J. Pomeranchuk, {\it
Dokl. Akad. Nauk. SSSR} {\bf 92} (1953) 535, {\bf 92} (1953) 735;
A.B. Migdal, {\it Phys. Rev.} {\bf 103} (1956) 1811;
R. Blankenbecler and S.D. Drell, {\it Phys. Rev. D}
{\bf 53} (1996) 6265; R. Baier, Yu.L. Dokshitzer,
A.H. Mueller, S. Peign\'e, and D. Schiff, hep-ph/9604327;
 P.L. Anthony et al., {\it Phys. Rev Lett.}
{\bf 75} (1995) 1949; SLAC-PUB-7413 (February, 1997).
\bibitem{bourbaki} N. Bourbaki, {\it Lie Groups and Lie Algebras},
(Hermann, 1975).

\end{thebibliography}
\end{document}